\documentclass[prd,floatfix,eqsecnum,fleqn,tightenlines]{revtex4} 

\usepackage{amsmath,amssymb,revsymb,graphicx,dcolumn}

\begin{document}

\title{Flat bands in topological media\vspace*{5mm}}

 \vspace*{5mm}\noindent

\begin{abstract}
  Topological media are systems whose properties are protected by
  topology and thus are robust to deformations of the system.  In
  topological insulators and superconductors the bulk-surface and
  bulk-vortex correspondence gives rise to the gapless Weyl, Dirac or
  Majorana fermions on the surface of the system and inside vortex
  cores. Here we show that in gapless topological media, the
  bulk-surface and bulk-vortex correspondence is more effective: it
  produces topologically protected gapless fermions without dispersion
  -- the flat band. Fermion zero modes forming the flat band are
  localized on the surface of topological media
  with protected nodal lines\cite{SchnyderRyu2010,HeikkilaVolovik2010b} and in the vortex core
  in systems with topologically protected Fermi points (Weyl points) \cite{Volovik2010}. Flat band has an extremely singular density of
  states, and we show that this property may give rise in particular
  to surface superconductivity which could exist even at room temperature.
 \end{abstract}


\author{Tero T. Heikkil\"a}
\email{tero.heikkila@tkk.fi}
\affiliation{Low Temperature Laboratory, Aalto University, P.O. Box 15100, FI-00076 AALTO, Finland}
 \author{N.B. Kopnin}
\email{kopnin@boojum.hut.fi}
\author{G.E. Volovik}
\email{volovik@boojum.hut.fi}
\affiliation{Low Temperature Laboratory, Aalto University, P.O. Box 15100, FI-00076 AALTO, Finland\\
and\\
L.D. Landau Institute for Theoretical Physics, Kosygina 2, 119334 Moscow, Russia}

\maketitle

\section{Introduction}
Topological matter is characterized by a nontrivial topology  in momentum space\cite{Volovik2003,Horava2005,HasanKane2010,QiZhang2010}.
This topology may be represented by the momentum-space invariants as depicted in Fig.~1. They are in many respects similar to the real-space invariants which describe topological defects in condensed matter systems and in particle physics. In particular, the Fermi surface in metals is topologically stable, because it is analogous to the vortex loop in superfluids or superconductors\cite{Volovik2003}. In the same way, the Fermi point (the Weyl point) corresponds to the real-space point defects, such as hedgehog  in ferromagnets or magnetic monopole in particle physics. The fully gapped topological matter, such as topological insulators and fully gapped topological superfluids represent skyrmions in momentum space: they have no nodes in their spectrum or any other singularities, and they correspond  to non-singular objects in real space -- textures or skyrmions (Fig.~1 {\it top right}).

Recently the interest to topological media has been mainly concentrated on the fully gapped  topological media, such as topological insulators and superfluids or superconductors of the $^3$He-B type. These systems contain topologically protected gapless fermions on the surface\cite{HasanKane2010,QiZhang2010},
and in the core of topological objects\cite{TeoKane2010,SilaevVolovik2010,FukuiFujiwara2010}, some of which have an     exotic Majorana nature.

The gapless topological media also exhibit exotic fermion zero modes with interesting properties. In particular they may have a Fermi arc (Fermi surface which terminates on a monopole in Fig.~1 {\it bottom right})) \cite{Tsutsumi2011,XiangangWan2011,Burkov2011}, and
a dispersionless branch of the spectrum with zero energy -- the flat band \cite{SchnyderRyu2010,HeikkilaVolovik2010b,Volovik2010,Sato2011,KopninHeikkilaVolovik2011,Kopnin2011}.
Historically  the flat bands were first discussed in relation to Landau levels, but they may emerge even without magnetic fields.   They were suggested in strongly interacting systems \cite{Khodel1990,NewClass,Volovik2007,Shaginyan2010}, in the core of quantized vortices\cite{KopninSalomaa1991}, in 2+1 dimensional quantum field theory which is dual to a gravitational theory in the anti-de Sitter  background\cite{Sung-SikLee2009}, in rhombohedral graphite\cite{Guinea2006} and on graphene edge \cite{Ryu2002}, and on the surface of superconductors with gap nodes in the bulk \cite{Ryu2002,SchnyderRyu2010,Sato2011}.  Flat band is the momentum-space analog of a domain wall (soliton) terminating on a half-quantum vortex\cite{NewClass} (Fig. 1 {\it bottom left}).
The topologically protected flat bands, which we discuss here, as well as the Fermi arc add a new twist in the investigation of the 3-dimensional topological matter, shifting the interest from the topological insulators and fully gapped superfluids/superconductors to their gapless 3-dimensional counterparts, such as superfluid $^3$He-A,  graphite, topological semi-metals and gapless topological superconductors.
Dispersionless bands may serve as a good starting point for obtaining interesting correlated and symmetry breaking states\cite{Green2010}.

This paper is based on the earlier arXiv version \cite{arXiv}.

\begin{figure*}
\includegraphics[width=\textwidth]{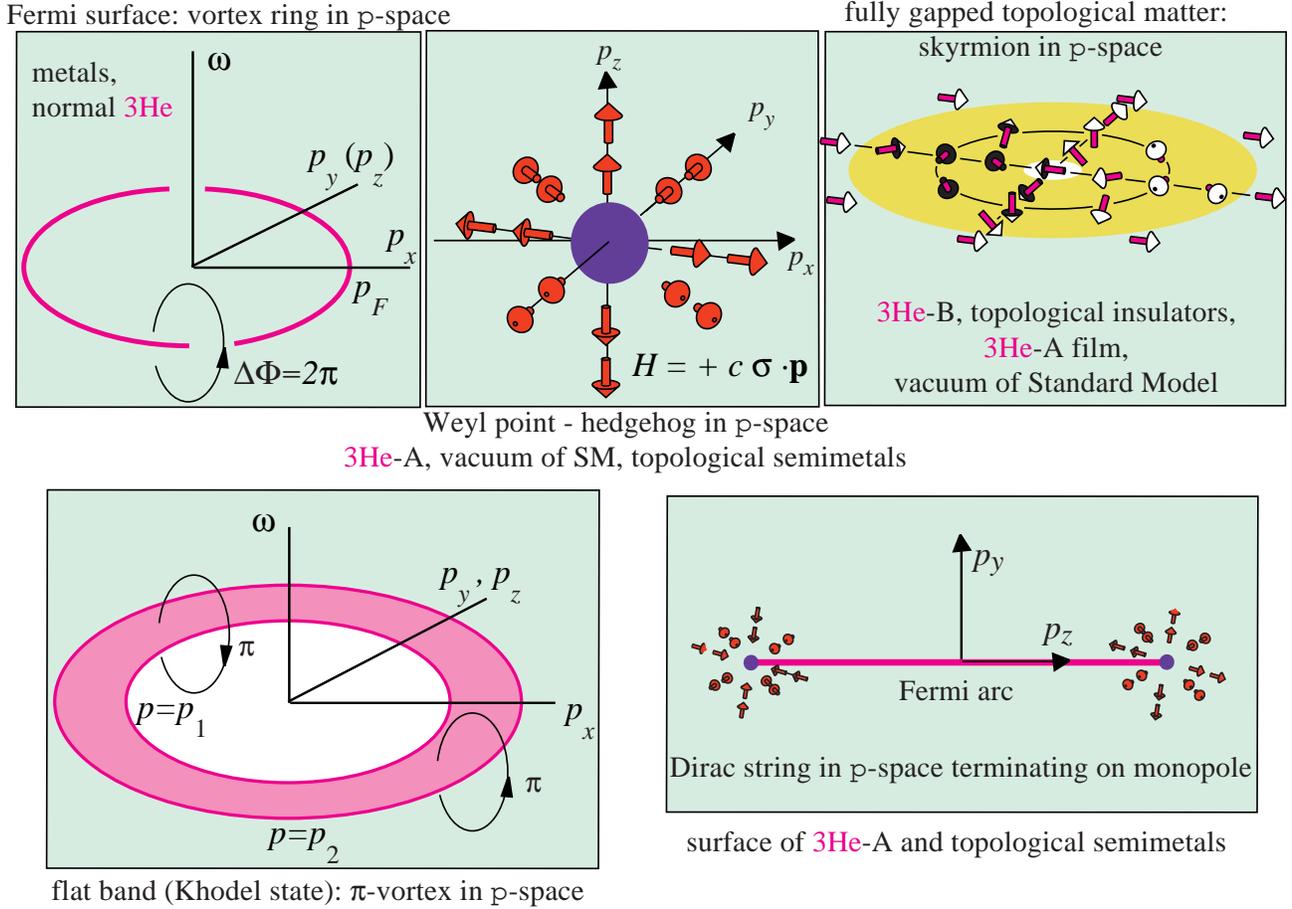}
\caption{Topological matter, represented in terms of topological objects in momentum space.
({\it top left}): Fermi surface is the momentum-space analog of the vortex line: the phase of the Green's function changes by $2\pi N_1$ around the element of the line in $(\omega, {\mathbf p})$-space. ({\it top middle}):  Fermi point (Weyl point) is the counterpart of a hedgehog and a magnetic monopole. The hedgehog in this figure has integer topological charge $N_3=+1$, and close to this Fermi point the fermionic quasiparticles behave as Weyl fermions. Nontrivial topological charges $N_1$ and $N_3$ in terms of Green's functions support the stability of the Fermi surfaces and Weyl points with respect to perturbations including interactions \cite{Volovik2003,EssinGurarie2011}. ({\it top right}):  Topological insulators and fully gapped topological superfluids/superconductors are textures in momentum space: they have no singularities in the Green's function and thus no nodes in the energy spectrum in the bulk. This figure shows a skyrmion in the two-dimensional momentum space, which characterizes two-dimensional topological insulators exhibiting intrinsic quantum Hall or spin-Hall effect. ({\it bottom left}):  Flat band emerging in strongly interacting systems  \cite{Khodel1990}. This dispersionless Fermi band is analogous to a soliton terminated by half-quantum vortices: the phase of the Green's function changes by $\pi$ around the edge of the flat  band   \cite{NewClass}.  
({\it bottom right}): Fermi arc on the surface of $^3$He-A \cite{Tsutsumi2011} and of topological semi-metals with Weyl points \cite{XiangangWan2011,Burkov2011}  serves as the momentum-space analog of a Dirac string terminating on a monopole. The Fermi surface formed by the surface bound states terminates on the points where the spectrum of zero energy states merge with the continuous spectrum in the bulk, i.e. with the Weyl points.}
\label{TopUnClasses}
\end{figure*}

\begin{figure*}
\includegraphics[width=\textwidth]{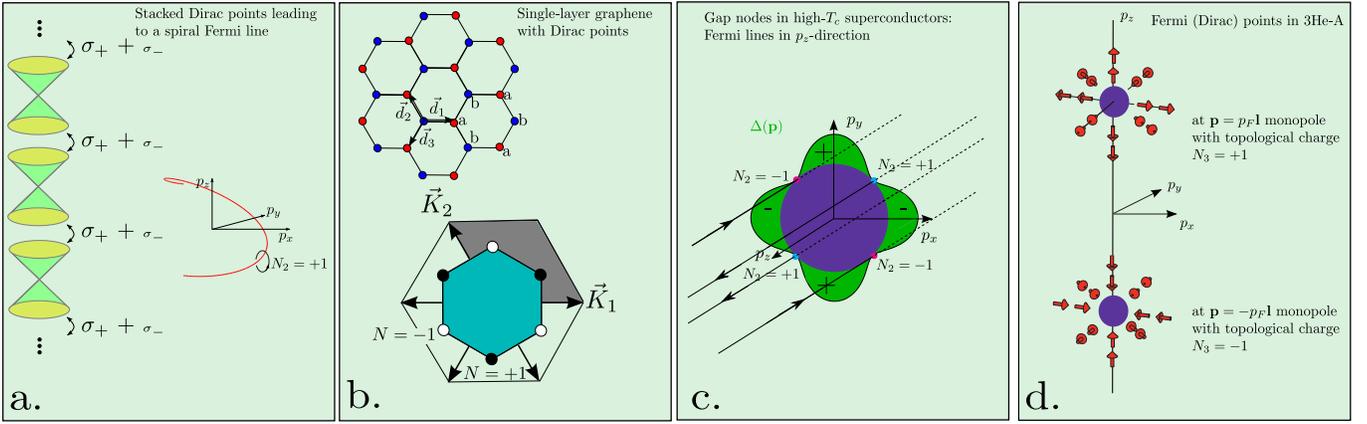}
\caption{Topological media which exhibit topologically protected dispersionless fermionic spectrum with exactly zero energy -- the flat band. ({\it a}): Three-dimensional topological semi-metal. Flat band can be obtained by stacking of graphene layers, each alone represented by Dirac points. In case of proper stacking, represented by the coupling of the pseudospins in each layer, the nodal line is formed in the bulk in the form of a spiral which ends up on the faces of the Brillouin zone\cite{HeikkilaVolovik2010b,McClure1969}.   This nodal line has a non-zero topological charge $N_2=1$.   This nontrivial charge protects the surface states with zero energy
$\epsilon(p_x,p_y)=0$, in the whole region within the projection of the spiral on the surface, see Fig.~3a.
({\it b}): Dirac points in graphene ({\it top: real-space crystal lattice; bottom: reciprocal space}). They lead to the flat band on the zig-zag edge, see Fig.~3b.
({\it c}): Cuprate superconductors have also topologically protected nodal lines\cite{Volovik2007,Beri2010}. They lead to the flat band on a side surface, see Fig.~3c. ({\it d}): Superfluid $^3$He-A has two three-dimensional Weyl points, with $N_3=+1$ and $N_3=-1$. The projections of the Weyl points on the direction of the vortex line determine the boundaries of the region where the spectrum of fermions bound to the vortex core is exactly zero, $\epsilon(p_z)=0$,  see Fig.~3d and Fig.~5.
}
\label{topologicalbulk}
\end{figure*}


\begin{figure*}
\centering
\includegraphics[width=\textwidth]{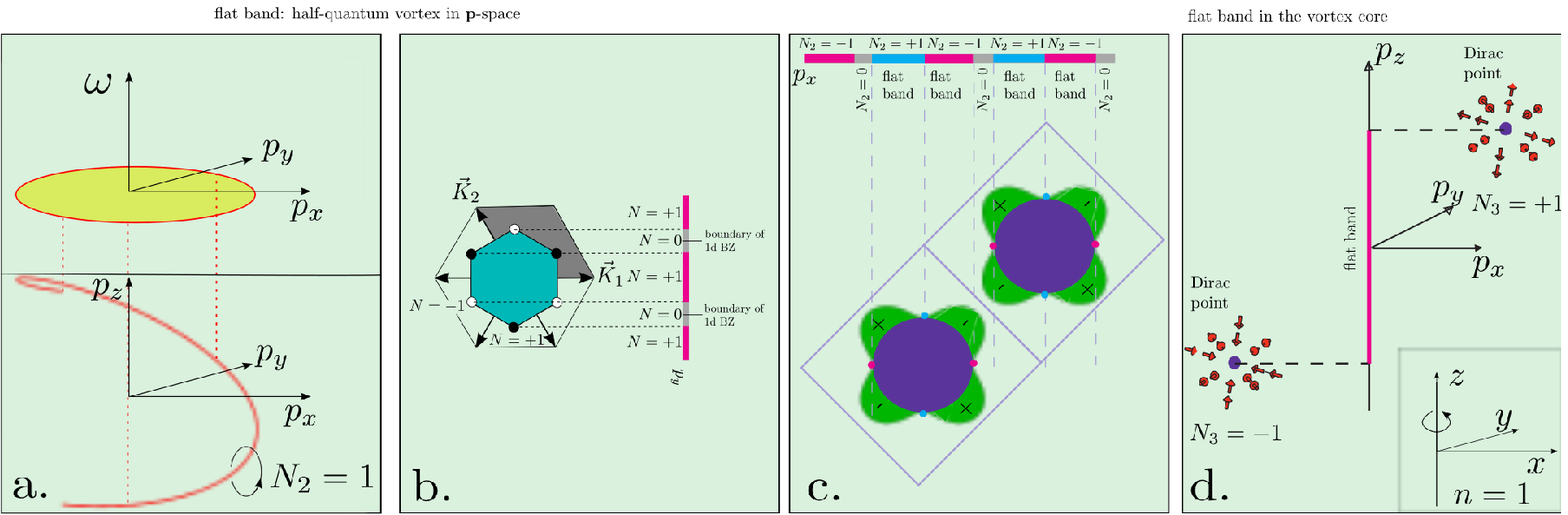}
\caption{Flat band formation.
({\it a, top}):    Two-dimensional flat band appears on the surface  of gapless systems with a topologically protected nodal line\cite{SchnyderRyu2010,HeikkilaVolovik2010b}. ({\it a, bottom}):  The nodal spiral in a semi-metal. This nodal line has a non-zero topological charge in momentum space, $N_2=1$, and this charge protects the surface states with zero energy
$\epsilon(p_x,p_y)=0$, in the whole region within the projection of the spiral on the surface.
({\it b}):  One-dimensional topologically protected flat band emerges on the zig-zag edge of graphene.
Projections of the Dirac points on the edge determine the boundaries of the flat band
\cite{Ryu2002}.
({\it c}): Nodes (nodal lines)  in cuprate superconductors give rise to the two-dimensional flat bands on the lateral surface
\cite{Ryu2002}. Projections of the nodes on the surface determine the boundaries of the flat bands.  ({\it d}): One-dimensional topologically protected flat band emerges in the core
of linear topological defects (such as the vortex with winding number $n=1$ in real space in the {\it bottom right corner}) in three-dimensional topological matter with Weyl points. The projections of the Weyl points on the direction of the vortex line  (along the $z$-axis) determine the boundaries of the region where the spectrum of fermions bound to the vortex core is exactly zero, $\epsilon(p_z)=0$, see also Fig.5.
 }
\label{BulkCorrespondence}
\end{figure*}

 \section{Surface flat band in a semimetal}

Consider the semimetal with topologically protected nodes in the form of a spiral as in Figs.~2{\it a} and 3{\it a}. The topological invariant in the bulk, supporting the existence and topological stability of the nodal line and as a result of the flat band with respect to interactions, is the contour integral in momentum space
\begin{equation}
N_2=- {1\over 4\pi i} ~{\rm tr} ~\oint_C dl  \sigma_z H^{-1}\nabla_l H\,,
\label{InvariantForLine2}
\end{equation}
 where $H({\mathbf p})$ is the effective matrix Hamiltonian (inverse Green's function at zero energy), and $\sigma_z$ is pseudospin. The nodal line is stable if the integral over the contour $C$ around the line is nonzero.
 On the other hand one may choose the contour $C$ as a straight line along the direction ${\mathbf p}_z$ normal to the surface, i.e. at fixed momentum  ${\mathbf p}_\parallel=(p_x,p_y)$ along the surface.
 Due to periodic boundary conditions, the points $p_z=\pm \pi/a$ are equivalent and the contour of integration $C$ forms a closed loop, giving integer values to the integral  $N_2({\mathbf p}_\parallel)$ if the integration path does not cross the point in the bulk where the energy is zero.  This integral $N_2({\mathbf p}_\parallel) =1$ for any point
 ${\mathbf p}_\parallel$ within the projection $S_p$ of the spiral on the surface, and $N_2({\mathbf p}_\parallel) =0$ outside this region. The states with momentum ${\mathbf p}_\parallel \in S_p$ cannot be adiabatically transformed into states in topologically trivial ($N_2=0$) media, and therefore a surface state with zero energy is formed for all momenta within $S_p$.\cite{HeikkilaVolovik2010b,SchnyderRyu2010,Ryu2002} This is in contrast with topological insulators, where this type of momentum-space invariants can be defined only for some particular values of momenta (but in principle, the insulators with topologically protected  surface flat bands are possible, we are indebted to A. Kitaev for this comment).

Such surface band emerges for example in the system of the
rhombohedral stacking (123123\ldots) of graphene layers\cite{CastroNeto2009,HeikkilaVolovik2010b} in the limit of a large
number $M$ of layers. Simultaneously the nodal line is formed in
the bulk, which is the source of the topological protection of the
surface band. The robustness of the flat band to disorder,
represented by stacking faults, is demonstrated in Fig.~4{\it c}.
The flat band remains even if the nodal line is transformed
to the Fermi surface tube as in rhombohedral graphite\cite{McClure1969}, if the invariant as the integral
around the Fermi tube is conserved.

In an analogous way stacking layers of topological insulators leads to the formation of a semimetal with two Weyl points in the bulk
and a Fermi arc on the surface \cite{Burkov2011}. The same happens in the process of the growth of the thickness of a $^3$He-A film, when two Weyl points are formed in the bulk  \cite{Volovik1992}, leading to a Fermi arc on the surface \cite{Tsutsumi2011}.

\begin{figure*}
\includegraphics[width=\textwidth]{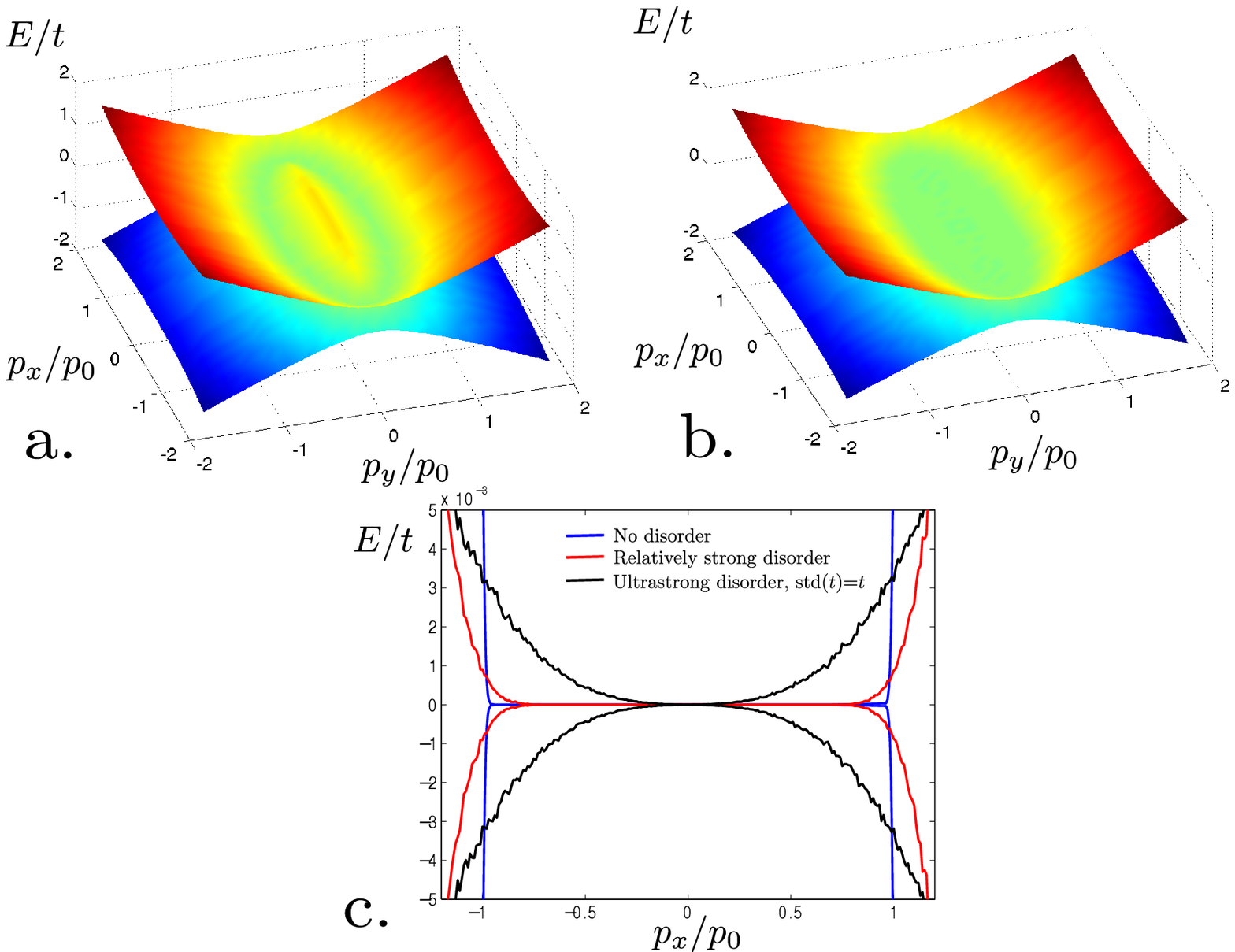}
\caption{Formation of the flat band and nodal line by stacking
$M$ graphene-like layers in the limit of large $M$. Flat band is
formed from the multiple Dirac point  on the surface, while the
nodal line is formed from the gapped states in the bulk. ({\it a}):
Formation of zero energy states in the bulk. Minima of the bulk energy
spectrum at fixed ${\mathbf p}_\parallel=(p_x,p_y)$ approach zero when $M\rightarrow
\infty$. These zeroes correspond to the projection of the spiral in
Fig.~3{\it a}  onto the surface. ({\it b}):
In the same limit $M\rightarrow \infty$ the surface flat band is
formed. Its boundary coincides with the projection of the nodal
spiral onto the surface. This demonstrates the bulk-surface
correspondence for the flat band. ({\it c}): Robustness of the surface
flat band to the stacking fault disorder.  The blue curves show
the flat band eigenvalues vs. $p_x$ for $p_y=0$ in the case when
there is only one coupling $t_+=t$ in the model \cite{HeikkilaVolovik2010b}. The red curves show the eigenvalues
in the case of Gaussian-distributed disorder in both $t_+$ and
$t_-$ couplings with a standard deviation equal to $0.5 t$ and the
black curves in the case when the disorder in both couplings
equals the average coupling $t$. In the previous case the flat
band survives, whereas in the latter case of strong disorder it is
destroyed. The results are obtained for $M=200$ layers and
averaged over disorder.} \label{Robustness}
\end{figure*}

 \section{Flat bands on the surface of cuprate superconductors and  graphene edge}

Dirac points in graphene and nodal lines in high-$T_c$ superconductors have also nontrivial
topological charges $N_2=\pm 1$ (see e.g. \cite{Volovik2007,Beri2010}). This results in the flat bands on the lateral surface of the
superconductor and on the special edges of graphene \cite{Ryu2002}. Figure 3{\it b} demonstrates the
1D flat band on the zig-zag edge of graphene and figure 3{\it c}  -- the 2D flat band on commensurate surface (1,1,0) of $d$-wave superconductor.
Projections of the nodes on the surface determine the boundaries of the regions
at which the integral $N_2({\mathbf p}_\parallel)$ takes a definite value: $0$,
$ +1$ or $-1$  (here ${\mathbf p}_\parallel=p_y$ is momentum along the graphene edge and ${\mathbf p}_\parallel=(p_x,p_z)$ for momentum along the surface of the superconductor; integration in (\ref{InvariantForLine2}) is now
along the momentum $p_\perp$ perpendicular to the surface/edge).
For any point ${\mathbf p}_\parallel$ for which $N_2({\mathbf p}_\parallel)\neq 0$, there is a
surface bound state with exactly zero energy $\epsilon({\mathbf p}_\parallel)=0$,
i.e. flat band.
For cuprate superconductors, this flat band corresponds to the infinite set of Andreev bound
states obtained
in the semiclassical approximation and
observed in grain-boundary junctions, see e.g.
\cite{ChiaRenHu1994,Alff1998,KashiwayaTanaka2000}. The topological analysis demonstrates that
even beyond the semiclassical approximation the Andreev
bound states form the flat band with exactly zero energy
in some regions of momentum space
if the surface has an orientation commensurate with the crystal. In this case the flat band consists of Majorana fermions, and for the considered
(1,1,0) surface there are boundaries which separate flat bands
with different topology, $N_2=+1$ and $N_2=-1$.

 \section{Flat band in a vortex core}

\begin{figure*}
\includegraphics[width=0.5\textwidth]{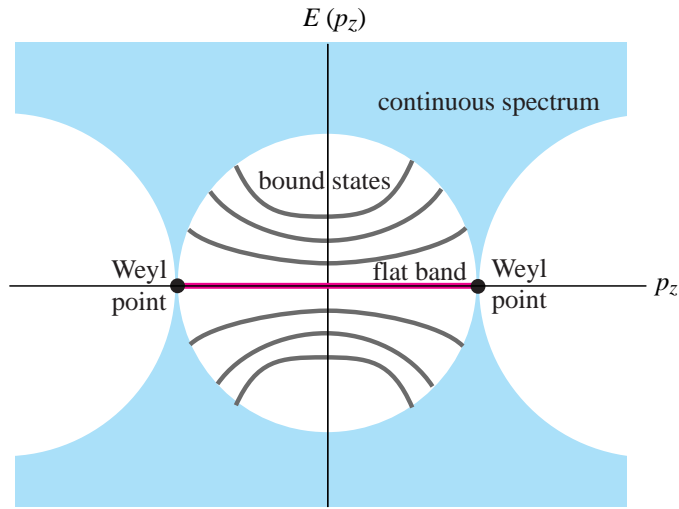}
\caption{Schematic illustration of the spectrum of bound states $E(p_z)$ in the vortex core. The branches of bound states terminate at points where their spectrum merges with the continuous spectrum in the bulk.
The flat band terminates at points where the spectrum has zeroes in the bulk, i.e. when it merges with Weyl points. It is another ${\bf p}$-space analog of a Dirac string terminating on a monopole, see Fig. 1 {\it bottom right} for the Fermi arc.} 
\label{VortexFlatBand}
\end{figure*}

The topologically protected  dispersionless spectrum with zero
energy appears also in linear topological defects -- vortices. The
topological protection is provided by the nontrivial topology of
three-dimensional Weyl points in the bulk. This bulk-vortex
correspondence\cite{Volovik2010} is illustrated in Fig.
3d for the case of a pair
of Weyl points with opposite topological charges $N_3=\pm 1$. The
projections of these Weyl points on the direction of the vortex
line determine the boundaries of the region where the spectrum of
fermions bound to the vortex core is exactly zero,
$E(p_z)=0$. Such flat band first obtained
in Ref.~\cite{KopninSalomaa1991} for the noninteracting model
is not destroyed by interactions. The spectrum of bound states 
in a singly quantized vortex in $^3$He-A is illustrated in Fig. \ref{VortexFlatBand}.

\section{Density of states, entropy and specific heat of a flat band}

The important consequence of the flat band is the anomalously singular density of states (DOS)
of electrons. The formation of a singular DOS with growing the number of layers
$M$ can be found in the model\cite{HeikkilaVolovik2010b}  in
which the spectrum of electrons in the surface layer contains the
multiple Dirac point with energy $\epsilon({\mathbf p}_\parallel)=
t (|{\mathbf p}_\parallel |/p_0)^M$, where $t$ is
the interlayer coupling.
For $M>2$ this system has a singular DOS $\nu(\epsilon)\sim
\epsilon^{2/M-1}$ and for $M \gg 2$ it is $\nu(\epsilon)=1/(2 \pi M\epsilon)$ at $\epsilon=0$.
Assuming for simplicity the spectrum $\epsilon({\mathbf p}_\parallel)=
t (|{\mathbf p}_\parallel |/p_0)^M$ for all $p<p_0$ one obtains the following entropy of the surface
flat band for large $M$
\begin{equation}
S\approx  \frac{AS_p}{(2\pi \hbar)^2 }  \left( 1-
\frac{2}{M} \ln \frac{t}{T}\right)\ln 2 ~~,~~ \frac{t}{T} \gg M\gg 2 \ln
\frac{t}{T}\,, \label{EntropyN}
\end{equation}
where $S_p=\pi p_0^2$ is the area of the flat band in
momentum space, and $A$ the area of the surface in real space. The
first term describing the flat band limit can be
understood as the entropy related to occupying or not occupying
the phase space volume $AS_p/(2\pi \hbar)^2$ of zero energy states. The second
term leads to a finite heat
capacity at low temperatures, $C_M(T)
\rightarrow AS_p\ln 2/(2\pi^2 \hbar^2M)$, which in the range of 
\eqref{EntropyN} dominates over the  bulk contribution.
The exact spectrum gives the same result which
corresponds to the equipartition law with the energy $T \ln 2 /(2M)$ 
per each degree of freedom in the flat band.

 \section{Superconductivity}

Singular DOS gives rise to superconductivity with an enhanced
transition temperature.
Already for $M=4$, $\nu(\epsilon)\sim \epsilon^{-1/2}$ corresponds to the DOS at
the edge of a one-dimensional band, which for a long time has been
discussed as a possible source of enhancement of the
superconducting transition temperature, see e.g. \cite{Kopaev1987}.
For general $M$ the self-consistency equation for the gap on the surface reads
\begin{equation}
\Delta =  \int\frac{V_p}{2} \frac{d^2 p}{(2\pi
\hbar)^2}\frac{\Delta}{E_{{\mathbf p}_\parallel}} \tanh\frac{E_{{\mathbf p}_\parallel}}{2T}
~~,~~E_{{\mathbf p}_\parallel}=\sqrt{\epsilon ^2({\mathbf p}_\parallel)+\Delta^2}
\ .  \label{surfaceSC}
\end{equation}
For simplicity we chose the pairing potential $V_{p}=g\theta(p_m-p)$ with $p_m\ll p_0$.
 For the flat band with $\epsilon ^2({\mathbf p}_\parallel)=0$ one obtains
\begin{equation}
\Delta =\frac{gp_m^2}{8\pi \hbar^2
}\tanh\frac{\Delta}{2T}\ .  \label{DeltaSC}
\end{equation}
For $T=0$ we get $\Delta(0)= gp_m^2/8\pi \hbar^2 $ and the
corresponding transition temperature is $T_c= \Delta(0)/2$. For finite $M$ one obtains
a power-law behavior of the superconducting gap at $T=0$,
\begin{equation}
 \Delta_S(0) = t\left(\frac{gp_m^2I_M}{8\pi
 \hbar^2 t}\right)^{\frac{M}{M-2}}
~~,~~I_M=\int_0^\infty \frac{ x^{\frac{M+2}{M}}\,
dx}{\sqrt{x^2+1}^3}\,. \label{SurfaceSCfiniteN}
\end{equation}
For $M\gg 1$ it approaches the linear law in Eq. (\ref{DeltaSC}).
For cut-off $p_m>p_0$ the more detailed calculation gives 
$\Delta(0)= gS_p/16\pi^2 \hbar^2$, 
where again $S_p$ is the area of flat band in momentum space. 
The
linear dependence of the gap and of the critical temperature on
the coupling should be contrasted with the exponential
suppression of $T_c$ in conventional bulk superconductivity, $T_c \sim
\exp(-1/\nu(0) g)$. The details can be found in \cite{KopninHeikkilaVolovik2011,Kopnin2011}.

The flat band should also emerge at the twin boundaries, if the
neighboring twins have opposite topological polarities, i.e.
opposite helicities of the nodal spiral. Such flat bands also support superconductivity, and thus a proper arrangement of many twins in the bulk would give bulk superconductivity with a large transition temperature.
Existence of localized superconducting domains at elevated
temperatures has been suggested for pure graphite and for
graphite-sulfur composites\cite{SilvaTorresKopelevich2001,LukyanchukKopelevich2009},
which may be related to the topology of nodes discussed here.

The strong dependence between the superconducting critical temperature and the coupling energy due to the singular density of states in the flat bands may open the route for room temperature superconductivity. However, also other types of broken symmetry states\cite{Bacsi2010}, such as magnetic instabilities\cite{Yazyev1990} or exciton BEC\cite{Basu2010} are possible, depending on the nature and magnitude of the corresponding coupling constants.

\section{Conclusion}

 Topological matter has generic behavior, depending on the topological class. Fermi surface, being topologically protected from interaction, results in the robustness of all the generic properties of metals at low temperatures. Systems with Fermi  points (Weyl points in 3D and Dirac points in 2D) represent other important classes of topological matter. Systems with Weyl points (superfluid $^3$He-A \cite{Volovik2003} and topological semimetals \cite{Abrikosov1971,Abrikosov1998,XiangangWan2011,Burkov2011}) experience emergent chiral or Majorana fermions and in addition the effective gauge fields and gravity as collective or artificial fields related to the deformations of the Weyl point. There are many reasons to believe that the quantum vacuum belongs to the class of topological media with Fermi points.  Quantum Hall effect, superfluid $^3$He-B and topological insulators have provided a new class of topological matter, which is characterized by topologically protected surface states and quantization of physical parameters, such as Hall and spin-Hall conductivity. The highly degenerate topologically protected state -- the flat band --  is also a generic phenomenon.  This new class of topological phenomena is waiting for its exploration.

\section*{\hspace*{-4.5mm}ACKNOWLEDGMENTS}
It is a pleasure to thank A.\ Geim, V.\ Khodel, A. Kitaev 
and K.\ Moler 
for helpful comments. This work
is supported in part by the Academy of Finland and its COE program
2006--2011, by the European Research Council (Grant No.
240362-Heattronics), by the Russian Foundation for Basic Research
(grant 09-02-00573-a), and by the Program ``Quantum Physics of
Condensed Matter'' of the Russian Academy of Sciences.


\end{document}